# Learning with Linear Mixed Model for Group Recommendation Systems

Baode Gao, Guangpeng Zhan, Hanzhang Wang, Yiming Wang, Shengxin Zhu
Department of Mathematics, Xi'an Jiaotong-Liverpool University
Suzhou 215123, Jiangsu Province, P.R. China
{Baode.Gao16, Guangpeng.Zhan15, Hanzhang.Wang16, Yimin.Wang16}@student.xjtlu.edu.cn, Shengxin.Zhu@xjtlu.edu.cn

## ABSTRACT

Accurate prediction of users' responses to items is one of the main aims of many computational advising applications. Examples include recommending movies, news articles, songs, jobs, clothes, books and so forth. Accurate prediction of inactive users' responses still remains a challenging problem for many applications. In this paper, we explore the linear mixed model in recommendation system. The recommendation process is naturally modelled as the mixed process between objective effects (fixed effects) and subjective effects (random effects). The latent association between the subjective effects and the users' responses can be mined through the restricted maximum likelihood method. It turns out the linear mixed models can collaborate items' attributes and users' characteristics naturally and effectively. While this model cannot produce the most precisely individual level personalized recommendation, it is relative fast and accurate for group (users)/class (items) recommendation. Numerical examples on GroupLens benchmark problems are presented to show the effectiveness of this method.

## 1. INTRODUCTION

Recommendation systems make a reasonable statistical inference or recommendation for business development through information collaborative filtering. A movie recommendation system is such an information filtering system. This kind of system aims to predict the preferences of users and comes up with suggestions. It is an active research area and has kept many researchers busy in the last decades. Recent experience has shown the success of matrix factorization methods based on sole rating data [6]. However, it is difficult to predict the inactive users' responses based on the limited rating information [11]. More information is needed to predict inactive user's response [11][1]. For instance, neural network based on embedding methods and collaborative filtering model have been used to model users' social connections (users may get purchasing advices from their friends and relatives) and purchasing habits. The effectiveness of these methods has been examined in [3]. Besides, trade-off between different preferences among group members is considered by a generative geo-social group recommendation model (GSGR), which provides analysis on the social relationship between team members and some specific preference of a certain group [4][13]. Alternatively, this paper will consider application of widely used linear mixed models (LMM) in genome wide association studies (GWAS) [14][7][9]. The application of recommender system in genome wide association study was recently studied in [9]. Actually, generalized liner models (GLM), commonly used GWAS models, have been applied in the job recommendation by LinkedIn. By introducing ID level coefficients in addition to global regression coefficients, job applications to potential employees increased by 20%- 40% [12]. In this paper, we shall examine the power of LMM in mining latent association between user features (e.g., age, gender, occupation) and item features. The mixed effects model will consider the objective factors as fixed effects and subjective factors as random effects. This model is valid to inactive users without rating history but with group characteristics (e.g. age, gender, occupation et. al) [8]. Stable GroupLens benchmark problems[4] are used to show the usefulness of proposed methods.

## 2. METHODOLOG: LINEAR MIXED MODELS

Linear mixed models are more accurate and realistic model than the linear regression models [10]. A mixed model consists of fixed effect terms and random effects terms. It allows hierarchical structures and leads to the so-called multi-level models. It can be written as

$$y = X\tau + Zu + \epsilon, \qquad (1)$$

where y is an $N \times 1$ vector of observations, $\tau$ is a $k \times 1$ column vector of fixed effects, $X$ is an $N \times k$ design/indicator matrix for the fixed effects, u is $j \times 1$ vector of random effects, $Z$ is $N \times j$ the design/indicator matrix for random effects. $\epsilon$ is random background noise. The random effects u and $\epsilon$ are supposed to be normal random variables with means zeros such that $E(u) = 0$, $E(\epsilon) = 0$, $var(u) = \sigma^2 G$, $var(\epsilon) = \sigma^2 R$ and

$$\text{var}\begin{pmatrix} u \\ \epsilon \end{pmatrix} = \sigma^2 \begin{pmatrix} G(\gamma) & 0 \\ 0 & R(\phi) \end{pmatrix}. \qquad (2)$$

where $G = G(\gamma)$ and $R = R(\phi)$ are parametric co-variance matrices. When G and R are known the fixed effects and random effects can be found by solving the mixed model equation

$$\begin{pmatrix} X^T R^{-1} X & X^T R^{-1} Z \\ Z^T R^{-1} X & X^T R^{-1} Z + G^{-1} \end{pmatrix} \begin{pmatrix} \hat{\tau} \\ \tilde{u} \end{pmatrix} = \begin{pmatrix} X^T R^{-1} y \\ Z^T R^{-1} y \end{pmatrix} \quad (3)$$

or by the Gauss-Markov-Aitken least square. It is known that such estimation $\hat{\tau}$ and $\tilde{u}$ are the Best Linear Unbiased Estimators (BLUEs). Let

$$C = \begin{pmatrix} X^T R^{-1} X & X^T R^{-1} Z \\ Z^T R^{-1} X & Z^T R^{-1} Z + G^{-1} \end{pmatrix} \quad (4)$$

It is known that confidence and uncertainty of estimations of random effects and random effects can be quantified by co-variance of the estimators and the predictors

$$var \begin{pmatrix} \hat{\tau} \\ \tilde{u} - u \end{pmatrix} = \sigma^2 C^{-1} \quad (5)$$

For the case when $\theta = (\sigma^2; \gamma; \phi)$ are unknown, we should first obtain a good variance parameter estimation according to a nonlinear iterative method for the restricted maximum likelihood method (REML) or the marginal likelihood method. Such a method is conceptually simple, though it is computationally demanded. The reader is redirected to [15-18] for more technique details. And there are several software packages available to estimate the fixed effects and random effects, for example, Python Statistics Models, R-limer4 packages[5]. Commercial software SAS, SPSS and Matlab also have robust procedures for the REML methods[2].

## 3. EXPERIMENT DESIGN

The successful application of the linear mixed model largely depends on proper experiment design and rigorously statistical test. We shall first demonstrate how the linear mixed model works for recommendation of a single movie to target user groups, and then we shall apply the model for group recommendation by associating users' characteristics with movie genre.

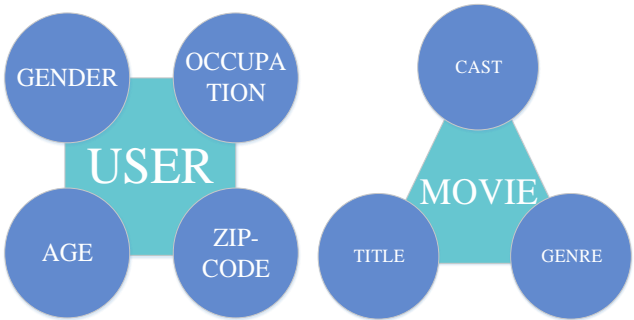

**Diagram 1. User and Item Information**

### 3.1 Recommendation of a Single Movie

In this model, we shall consider whether the response of rating of a given movie is impacted by user's characteristics (such like age, occupation, gender, region and so on). By such a simple model, one would answer the question like whether students prefer *Jurassic Park* than lawyer. Which characteristics results more various response? In this case, the response value can be modeled as

○ Model 1: $y \sim -1 + age + (1|occupation)$
○ Model 2: $y \sim -1 + occupation + (1|age)$
○ Model 3: $y \sim -1 + gender + (1|age)$
○ Model 4: $y \sim -1 + occupation + age + (1|gender)$

Here we use the patsy formula in R, Python and Matlab, where y is the response value. In Model 1, age groups are viewed as fixed effects, occupation and gender are viewed as random effects. The -1 terms in the models indicates that this is a model without random intercept terms. For the GroupLens dataset, the users are divided to 21 occupations and are belong to 7 age groups (see appendix for more information). More specifically, the design matrix X for the fixed effects $\tau$ is an $n \times 21$ matrix. Each row of X contains only one nonzero element 1 which indicates the underlying user's occupation. Other elements are zeros. Z is a $n \times 9$ design matrix of random effects, where there are 7 columns indicates the age group and 2 columns represents the gender.

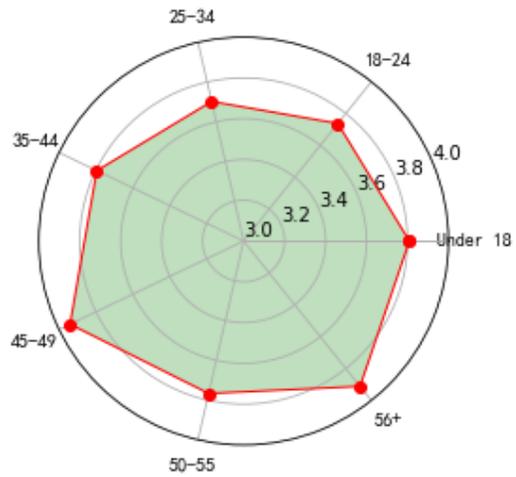

**Figure 1. Model 1's Fixed Effects Coefficients for Jurassic Park**

Fig.1 and Fig.2 illustrate the fixed effects of model 1 and model 2, which are the average scores of a given group to the underlying movie. As shown in Fig.2, occupation used as fixed effects results in significant response values compared with age groups as fixed effects. More rigorously, we can use statistical inference methods to infer which factor play a more significant effect. For example, this can be done by computing the ANOVA p-value of the fixed effects. The main statistical criteria, P-value, Akaike information criterion (AIC), Bayesian information criterion (BIC), and log likelihood (logL), are listed in Table 1. For Model 2 and Model 3, the p-values are very small which indicate that occupation and gender can be viewed as fixed effects. And the coefficients of the fixed effects reveal the preference of different groups to the movie *Jurassic Park*. For example, for Model thee the fixed coefficient for male is 3.8142 and for female is 3.5794, this shows that the Male more enjoy the *Jurassic Park* Movie.

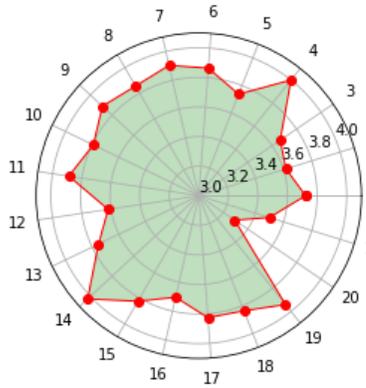

**Figure 1. Model 2's Fixed Effects Coefficients for Jurassic Park**

**Table 1. Statistical Tests**

|         | Model 1 | Model 2 | Model 3 |
|---------|---------|---------|---------|
| p-value | 0.0395  | 0.006   | 2.6e-7  |
| AIC     | 7569.9  | 7522.3  | 7409.7  |
| BIC     | 7617    | 7651.7  | 7433.3  |
| LogL    | -3777   | -3739.1 | -3700.9 |

To compare two nested models such as the model 2 and model 4, where the fixed effects of model 2 is a sub-factor of model 4, we can use the likelihood ratio test to compare the two model[5]. For example, in Matlab we can use the function compare (model1, model2), which returns some criteria as in Table 2. The results show that the hierarchal model, Model 4, is better than Model 2. In general, a hierarchal model or multilevel level is better.

**Table 2. Statistical Tests for Model 2 and Model 4**

|          | Model 2  | Model 4  |
|----------|----------|----------|
| DF       | 22       | 28       |
| AIC      | 7522.3   | 7483.8   |
| BIC      | 7651.7   | 7648.5   |
| LogL     | -3739.1  | -3713.9  |
| LRStat   | --------- | 50.465   |
| Delta DF | --------- | 6        |
| P- value | --------- | 3.7e-09  |

## 3.2 Recommendation of a Movie Genre

In a more realistic application, we want to associate user groups to movie genre so that we can promote a class of movie to certain target groups. We shall confirm that the response to a type of movie may vary from group to group. By studying this problem, the question like which age group people prefer a certain type of movie can be solved. We use the similar technic and model design as section A. However, instead consider only one movie, we shall consider a type of movie.

○ Model 5: $y \sim -1 + age + (1|occupation) + (1|gender)$
○ Model 6: $y \sim -1 + occupation + (1|age) + (1|gender)$

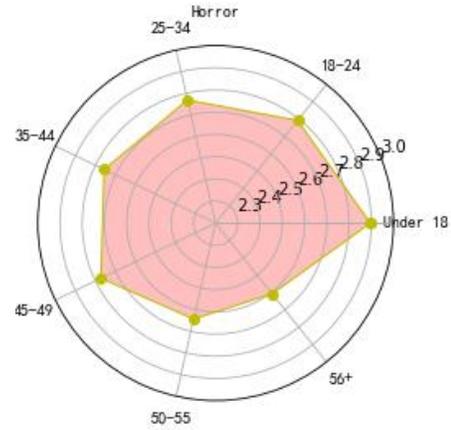

**Figure 2. Coefficients of Fixed Effect (Model 5) for Horror Movie**

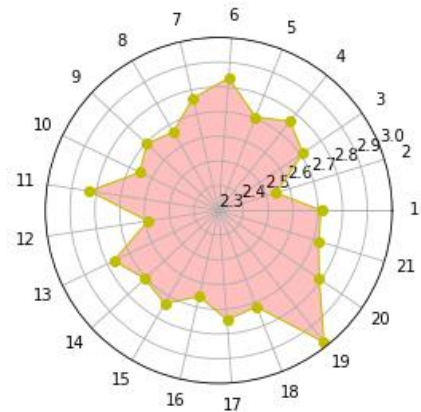

**Figure 3. Coefficients of Fixed Effect (Model 6) for Horror Movies**

Fig. 3 and Fig. 4 show the rating differences in virtue of different age and occupations are obvious. In age attribute, as shown in Fig.3, teenagers and the youth prefer horror movie. In occupation characteristic, Fig. 4 shows that the response of rating of craftsman is the highest for horror movies.

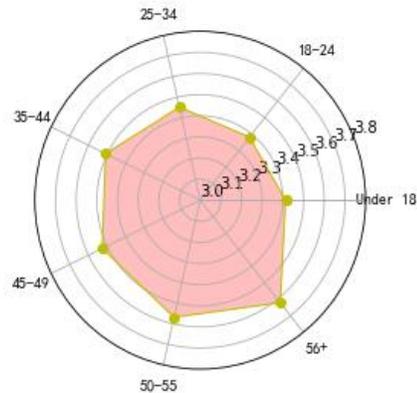

**Figure 4. Coefficients of Fixed Effect (Model 5) for Musical Movies**

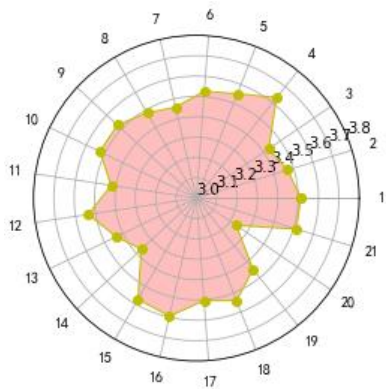

**Figure 5. Coefficients of Fixed Effect (Model 6) for Musical Movies**

As shown in Fig. 5, and Fig. 6, the artist responses the highest rating for musical movies and overall users' response of musical movies rating increase with age. This suggests that artist and the elder prefer elegant art movies and the artistic appreciation increases as a person grows.

## 3.3 Accurany of Prediction with LMM

In order to determine accuracy of the model, we use the commonly used indicator Mean Absolute Error (MAE) in recommendation systems. The MAE is defined by the average between the predict value the predicted value $P_i$ and actual value $a_i$.

$$\text{MAE} = \frac{\sum_1^n |P_i - a_i|}{n} \quad (6)$$

The less MAE, the more accurate prediction. We use the cross-validation method to compute the MAE indicator.

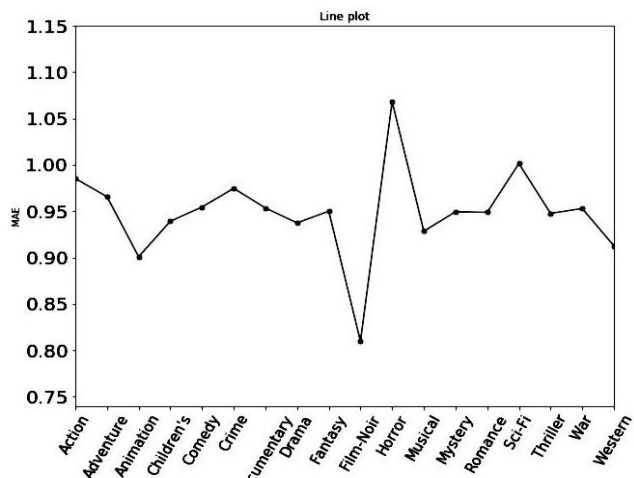

**Figure 6. MAE Values of Model 6 with Different Movie Genres**

We randomly select 80% data as training set and reserve 20% data as testing set. Once the parameters τ and u have been learned, we can predict the rating values for the testing data and then compare the accuracy between different models.

Here we compare our linear mixed model's MAE with the methods used in[1]. We use the same data as they did. From Fig. 7 and Fig.8, we can see that the linear mixed model produces much more smaller MAE than that method used in [1], and thus more accurate.

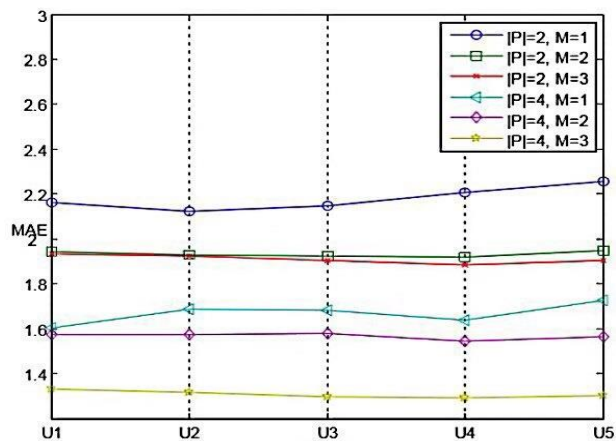

**Figure 7. MAE Values of Fusion Heterogeneous Information Network and Rating Matrix Reconciliation Method [1]**

## 4. CONCLUSIONS

In this paper, we explored the linear mixed model methods in recommendation systems. The model can find the latent association between users' characteristics and items' attributes through rigorous statistical test. The choice of fixed effects and random effects can be assumed and then verified by ANOVA test. Nested models with fixed effect can be compared. Such models are particularly effective in recommendation for new users or inactive users without rating history, provided some information of the users (like age, gender, occupation and region) can be obtained. Examples have showed some movies are popular to certain group of people, and some group of people prefer certain genre of movie. Therefore, these methods will be great interest for group recommendations and recommendations for new users or inactive users without rating history. The linear mixed model enjoys significantly better prediction accuracy, compared with some fusion heterogeneous information network methods. Compared with the conventional simple regression models, the linear mixed models are more advanced to describe the objective and subjective factors in a recommendation process, and thus is more promising and deserves better research in the future.

## 5. APPENDIX

### 5.1 Data Description

The data we used are GroupLens 1M Data, which can be accessed at http://grouplens.org/dataset/movielens/1m/. The classifications of age groups and occupations are as follows

**Table 3. User information**

| Occupations | Age |
|---|---:|
| 1: Other or not specified | Under 18 |
| 2: Academic/educator | 18-24 |
| 3: Artist | 25-34 |
| 4: Clerical/admin | 35-44 |
| 5: College/grad student | 45-49 |
| 6: Customer service | 50-55 |
| 7: Doctor/health care | 56+ |

| 8: Executive/managerial | Age |
|---|---|
| 9: Farmer | Under 18 |
| 10: Homemaker | 18-24 |
| 11: K-12 student | |
| 12: Lawyer | |
| 13: Programmer | |
| 14: Retired | |
| 15: Sales/marketing | |
| 16: Scientist | |
| 17: Self-employed | |
| 18: Technician/engineer | |
| 19: Tradesman/craftsman | |
| 20: Unemployed | |
| 21: Writer | |

## 5.2 Data Processing

There were 1,000,209 anonymous ratings of approximately 3,900 movies made by 6,040 GroupLens users who joined GroupLens in 2000. In order to solve cold start problem (in accurate recommendation for new users' or inactive users' movies), we determined age, occupation and gender as factors in linear mixed models. And we fit models for a type of movies of the same genre. For items' attributes, we classify movies by genres. We put every multi-label movie data into all its subtype data sets. This is the simplest way to solve multi-labeled items for linear mixed models.

## 6. ACKNOWLEDGMENT

The research is supported by National Natural Science Foundation (No. 11501044), Natural Science Foundation of Jiangsu Province (BK20171237) and research development found of Xi'an Jiaotong-Liverpool University (RDF-2017-02-23)